\documentclass[prl,twocolumn,showpacs]{revtex4}
\usepackage{graphicx}
\usepackage{amsmath}
\usepackage{amsfonts}
\usepackage{amssymb}
\usepackage{epsfig}
\usepackage{color}

\begin{document}

\title{Large ion Coulomb crystals: a near-ideal medium for coupling optical cavity modes to matter}

\author{A. Dantan, M. Albert, J. P. Marler, P. F. Herskind and M. Drewsen}

\affiliation{QUANTOP, Danish National Research Foundation Center for Quantum Optics, Department of
Physics and Astronomy, University of Aarhus, DK-8000 \AA rhus C., Denmark}

\begin{abstract}
We present an investigation of the coherent coupling of various transverse field modes of an optical cavity to ion Coulomb crystals.
The obtained experimental results, which include the demonstration of identical collective coupling rates for different transverse modes of a cavity field to ions in the same large Coulomb crystal, are in excellent agreement with theoretical predictions. The results furthermore suggest that Coulomb crystals in the future may serve as near-ideal media for high-fidelity multi-mode quantum information processing and communication purposes, including the generation and storage of single photon qubits encoded in different transverse modes.
\end{abstract}


\date{\today}

\maketitle

For the field of quantum information to mature to a practical stage, where complex computational tasks~\cite{nielsen_quantum_2000} and quantum information networks~\cite{kimble08} can be realized, careful attention has to be paid to the actual quality of the physical systems involved.  While the ideal system for carrying out serious quantum computational tasks has yet to be singled out, photons remain uncontested in their position as the ideal carriers of quantum information over large distances. Consequently, there is a clear need for identifying high quality physical media suitable for e.g. single photon generation and storage as well as for the realization of quantum repeaters. 

Among the systems under consideration (e.g systems discussed in Ref.~\cite{kimble08,lounis05}), a single ion in an optical cavity appears to be an excellent candidate for such realizations~\cite{guthohrlein01,keller2004,kreuter2004,leibrandt}. In addition, larger ensembles of cold trapped ions in the form of so-called Coulomb crystals~\cite{wineland87,walther92,drewsen98} in cavities may afford new opportunities for not only single-, but also multi-mode quantum information purposes, due to the combination of their solid state properties, including long term stability and uniform ion density throughout the crystal, and their single isolated particle features, i.e. no significant internal state perturbations due to interactions between the individual ions~\cite{herskind09NatPhys}. These features are relevant for e.g. encoding quantum information in orthogonal transverse spatial modes in contrast to the usual application of frequency and polarization degrees of freedom ~\cite{vasilyev08,andersen06}. In addition, coupling to multiple modes may have applications even outside of the context of quantum information, for e.g. quantum imaging~\cite{kolobov00,boyer08} and cavity mediated cooling~\cite{vuletic01,horak01,kruse03}.

Important steps towards the realization of high quality single-mode quantum devices based on an ion Coulomb crystal in a cavity were very recently taken, by demonstrating that the collective strong coupling regime of Cavity Quantum ElectroDynamics~\cite{haroche_exploringquantum_2006} can be reached with large ion Coulomb crystals interacting with a cavity field at the single photon level in the fundamental TEM$_{00} $ mode of an optical cavity~\cite{herskind09NatPhys}.

In this paper, the potential of ion Coulomb crystals as a media for multi-mode quantum interfaces is explored by a thorough experimental investigation characterizing the coupling of the TEM$_{00}$ and TEM$_{10,01}$ transverse cavity-mode fields at the single photon level to ion Coulomb crystals of various dimensions. 

Using needle shaped crystals aligned along the cavity axis with a diameter smaller than the waist characterizing the Hermite-Gaussian TEM$_{mn}$ field modes, detailed information on the coupling strength can be obtained through changing the position of the Coulomb crystal in the plane perpendicular to the cavity axis. Since through this procedure the crystal is moved into regions of dramatically varying micromotion induced by the trapping fields, the influence of this quiver motion on the coupling strength can be judged. In complementary experiments, the scaling of the coupling strength with respect to the crystals radial size for different modes has been investigated using Coulomb crystals with the central axis kept fixed in position and length in order to determine if and when convergence of the coupling strengths is reached. Finally, a precise measure of the coherent coupling rate for both modes is obtained in the regime where the coupling is expected to be independent of the cavity field mode (i.e. the radius of the crystal is large compared to the cavity mode waist), and additionally the number of ions is sufficient to attain useful optical densities.  Throughout, comparison with theory is made to find near-perfect agreement in all instances.

\begin{figure}
  \includegraphics[width=7.5 cm]{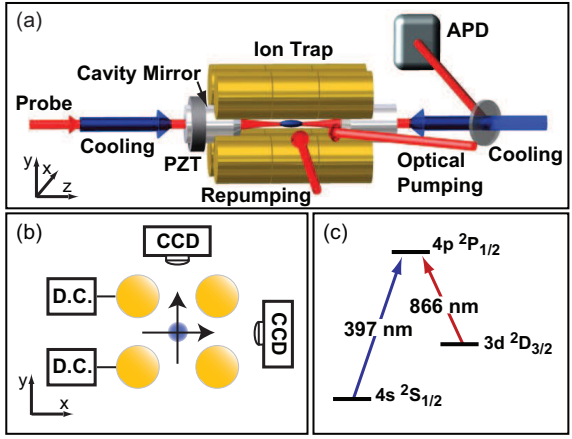}
  \caption{(color online.) (a)Schematic diagram of the experimental set-up with the linear rf trap
incorporating an optical cavity along its rf-field-free axis ($z$-axis). 
One of the cavity mirrors is mounted on a piezoelectric transducer (PZT) allowing for tuning of the cavity resonance around the 3d $^2$D$_{3/2}-$ 4p $^2$P$_{1/2}$ transition of $^{40}${Ca}$^+$. Also shown is the location and direction of the laser beams used in the experiment. (b) End-view schematic: The crystals can be displaced in the ($xy)-$plane by application of appropriate dc voltages to two of the segmented electrode rods, their position and size is monitored using two CCD cameras. (c) Relevant energy levels for the $^{40}$Ca$^+$ ion.} \label{fig1}
\end{figure}

In Fig.~\ref{fig1}a schematic of the experimental setup is shown. The
trap used is a linear rf trap with an optical cavity
incorporated in between the four segmented electrode rods~\cite{herskind08}, such that the cavity axis is parallel to the
trap's rf field-free axis ($z$-axis). $^{40}$Ca$^+$ ions are loaded
into the trap via resonant two-photon photoionization~\cite{herskind08} and subsequently Doppler cooled on the 4s
$^2$S$_{1/2}$-4p $^2$P$_{1/2}$ transition using two
counter-propagating 397 nm laser beams while repumped on the 3d
$^2$D$_{3/2}-$ 4p $^2$P$_{1/2}$ transition with a 866 nm laser beam~\cite{herskind09NatPhys}. After the ions are cooled into a crystalline state, they are prepared by optical pumping with an efficiency of $\sim 97\%$~ into the $m_J=+3/2$ magnetic sub-state of the long-lived metastable D$_{3/2}$ level \cite{herskind09NatPhys}. The optical cavity consists of two mirrors with a radius-of-curvature of 10 mm, which are separated by 11.8 mm. The cavity has a finesse of 3000 for light at the wavelength of 866 nm used to induce the coupling to the $^{40}$Ca$^+$ ions via the 3d $^2$D$_{3/2}-$ 4p $^2$P$_{1/2}$ transition. The Hermite-Gaussian TEM$_{mn}$ modes at this wavelength have a waist of $w_0=37$ $\mu$m and a Rayleigh range $z_R\sim 5$ mm.

The coherent coupling is measured by injecting a weak probe laser pulse around 866 nm into the cavity, and detecting the transmitted light by an avalanche photodiode (APD) after spatial and spectral filtering.

The probe pulse is mode-matched to a given transverse mode of the cavity, and its intensity is chosen such that the mean photon number in the cavity is about or less than one at all times. The probe polarization is left-hand circularly-polarized in order to obtain the strongest interaction with the ions optically pumped into the $m_J=+3/2$ magnetic sub-state of the D$_{3/2}$ level.

In a typical experiment, the cavity is scanned 1.2 GHz
across the atomic resonance at a repetition rate of 30 Hz, while a sequence including
cooling (5 $\mu$s), optical pumping (12 $\mu$s) and probing (1.4
$\mu$s) is repeated at a 50 kHz rate. The cavity transmission spectrum is constructed by averaging
the signals obtained from typically a few hundred scans of the cavity.

In order to compensate for thermal drifts and mechanical vibrations of the cavity during the experiments, a stronger laser beam not interacting with the ions (wavelength 894~nm) is injected into the cavity and detected in transmission for reference.

For a certain frequency of the probe pulse, the transmission
spectrum of the scanned cavity is expected to have a Lorentzian shape,
but with a half-width dependent on the coherent
coupling to the ions in the Coulomb crystal. This half-width can formally be written as~\cite{herskind09NatPhys}
\begin{eqnarray}\label{kappa}
\kappa'=\kappa+G_{mn}^2\frac{\gamma}{\gamma^2+\Delta^2}\end{eqnarray}
where $G_{mn}$ is the collective coherent coupling rate with the
TEM$_{mn}$ mode considered, $\kappa=(2\pi)2.15$ MHz is the cavity
natural linewidth, $\gamma=(2\pi)11.2$ MHz is the optical dipole
decay rate of the 3d~$^2$D$_{3/2}-$4p~$^2$P$_{1/2}$ transition and
$\Delta$ is the detuning of the probe pulse with respect to the
atomic resonance. For ion Coulomb crystals in a linear rf quadrupole trap, which typically have a
spheroidal shape with an uniform density $\rho$ ~\cite{drewsen98}, the coherent
coupling rate is given by  

\begin{widetext}
\begin{eqnarray}\label{gmn} G_{mn}^2(x_0,y_0)=g^2\rho\int_{V}d{\bf r} \Psi_{m}^2(x-x_0,z)\Psi_{n}^2(y-y_0,z)\sin^2\left[kz-(m+n+1)\tan^{-1}\frac{z}{z_R}+k\frac{(x-x_0)^2+(y-y_0)^2}{2r(z)}\right]\end{eqnarray}
\end{widetext}
where $x_0$ and $y_0$ represent the radial offsets of the crystal
center axis with respect to the cavity mode symmetry axis, $g$ is
the single-ion coupling strength at an anti-node in the center of
the fundamental TEM$_{00}$ mode, $V=4\pi R^2L/3$ is the volume of the spheroidal crystal with half-length $L$ and radius $R$ and
$\Psi_{l}$ is the Hermite-Gaussian mode-function
\begin{eqnarray}\nonumber
\Psi_{l}(u,z)=\sqrt{\frac{w_0}{w(z)}}H_l\left(\frac{u\sqrt{2}}{w(z)}\right)\exp\left(-\frac{u^2}{w(z)^2}\right)\;\;(u=x,y)
\end{eqnarray} with $H_l$ being the $l$th Hermite polynomial,
$w(z)=w_0(1+z^2/z_R^2)^{1/2}$ and $r(z)=z+z_R^2/z$.

\begin{figure}
  \includegraphics[width=7 cm]{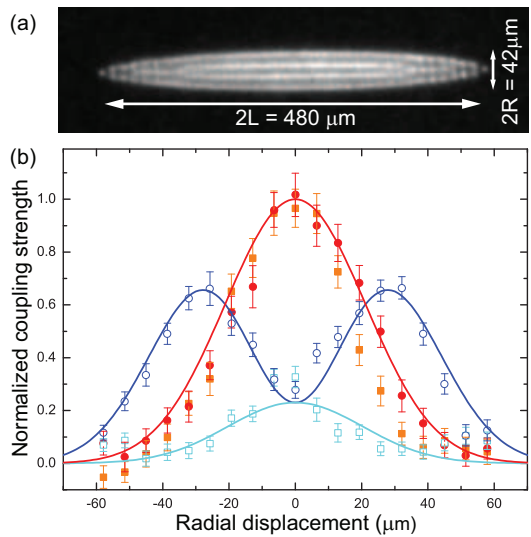}
\caption{(color online.) (a) Projection image in the $(xz)-$plane of the thin needle-shaped crystal used for the
data in Fig. 2b. (b) Normalized coherent coupling strengths $G_{00}^2$ (solid) and
$G_{10}^2$ (open) as a function of the displacement of the needle-shaped crystal along the $x$ (circle) and $y$ (square)-axes. The solid lines are derived from the theoretical expression given in
Eq.~(\ref{gmn}). } \label{fig2}
\end{figure}

Of particular relevance for evaluating the applicability of the system as a transverse multi-mode light-matter interface is the case of $R\gg w_0$, in which according to Eq.~2 the coupling strengths of the different transverse modes to the crystal are expected to converge.  However in order to achieve convergence, the effect of micromotion of the individual ions, including those located far from the trap axis, on the coupling strength must be negligible. To evaluate this effect, a thin needle-shaped
spheroidal Coulomb crystal, with half-length $L=240\pm 1$~$\mu$m and radius $R=21\pm 1$~$\mu$m (Fig.~2a), is
translated radially across the cavity mode-volume by the application
of appropriate dc voltages to two of the segmented electrode rods
(Fig.~1b). The displacement of the crystal in the transverse $(xy)-$plane is
determined by imaging the 397 nm fluorescence light emitted
during cooling onto two CCD cameras. (See sketch in Fig.~\ref{fig1}b.)
The precision in reading the position of the crystal is $\pm 0.8~\mu$m~\cite{herskind09JPhysB}. 

In Fig.~\ref{fig2}b the measured coupling strengths using the Coulomb crystal shown in Fig.~\ref{fig2}a is presented in terms of $G_{mn}^2$ normalized to $G_{00}^2(x_0=0,~y_0=0)$ for the TEM$_{00}$ and TEM$_{10}$ modes. For each position of the crystal, the coherent coupling strength is measured
through the broadening of the probe pulse transmission signal
(Eq.~\ref{kappa}) with the probe tuned to the atomic resonance ($\Delta=0$).
The experimental data is shown to agree very well with the
theoretical predictions of Eq.~(\ref{gmn}) with mode-functions derived from the cavity geometry, the crystal size and shape obtained from CCD images and the density of the ions determined from the trapping parameters~\cite{herskind09JPhysB}. Since the amplitude of the radial micromotion increases with the ions' distance from the rf-free $z$-axis, strong systematic deviations would have been expected at large displacements if micromotion was an issue.
This result importantly indicates that the inevitable radial micromotion of ions in linear rf traps does not couple significantly into their axial motion.

\begin{figure}
  \includegraphics[width=7 cm]{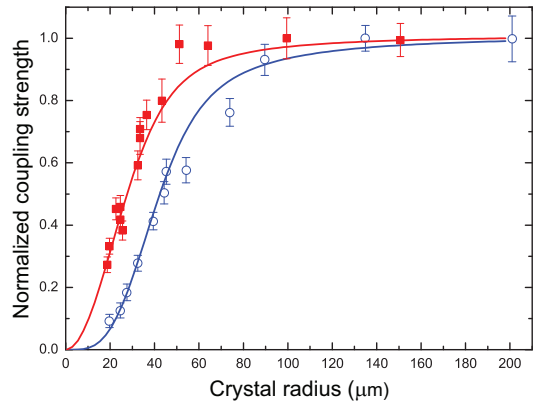}
  \caption{(color online.) Normalized coherent coupling strengths $G_{00}^2$ (solid)
  and $G_{10}^2$ (open) as a function of the crystal radius $R$, for crystals with a fixed length of $2L=672\pm 1\mu $m. The solid lines are derived from the theoretical expression given in Eq.~(\ref{gmn})} \label{fig3}
\end{figure}

To what extent ion Coulomb crystals constitute a near-ideal media with respect to coherent coupling to various cavity
modes has further been investigated by measuring the coherent coupling strength using ion Coulomb crystals with fixed length $2L=672\pm 1$ $\mu$m, but with varying radii. The resulting coherent coupling strengths for the TEM$_{00}$ and TEM$_{10}$ mode cases
are shown in Fig.~\ref{fig3}. The density of all the crystals is
$\rho=3.8 \times 10^{8}$ cm$^{-3}$, and their rotational symmetry
axes coincide with that of the cavity mode ($x_0=y_0=0$ in
Eq.~(\ref{gmn})). As expected, the increase in the coupling strength
as a function of the radius is slower for the TEM$_{10}$ than for
the TEM$_{00}$, since ions at the center of the cavity mode
contribute much less to the coupling for the former. However, the coupling
strengths converge to the same value for both modes for crystals
with large radii, in good agreement with the theoretical predictions
of Eq.~(\ref{gmn}).

\begin{figure}
  \includegraphics[width=7 cm]{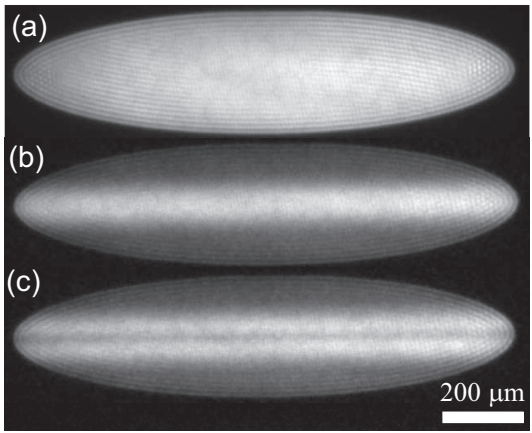}
  \caption{Projection images in the $(xz)-$plane of the $1.2$ mm-long Coulomb crystal used
  for the measurements in Fig. \ref{fig5}, when (a) the whole crystal is illuminated by 866~nm repumping light along the $x$-axis, (b,c) repumping light at 866~nm is predominantly injected into the (b) TEM$_{00}$ and (c) TEM$_{10}$ cavity modes, for enhancing the fluorescence level within these modes.} \label{fig4}
\end{figure}
\begin{figure}
  \includegraphics[width=7 cm]{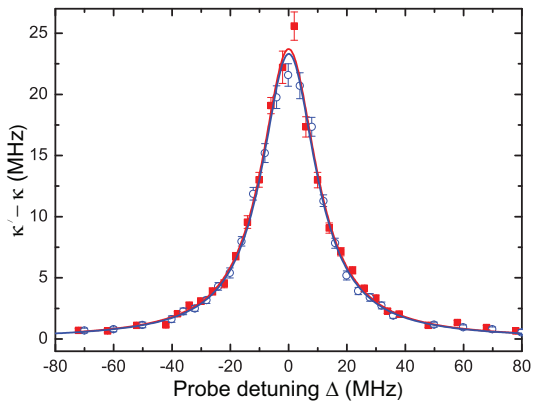}
  \caption{(color online.) Broadening of the probe signal half-width $\kappa'-\kappa$ as a function of the probe detuning $\Delta$, for the TEM$_{00}$ (solid) and the TEM$_{10}$ (open) modes, obtained with the crystal of Fig.~\ref{fig4}. The collective coupling rates are deduced from Lorentzian fits according to Eq.~\ref{kappa} (solid lines).} \label{fig5}
\end{figure}

Finally, to carefully check the prospect of using large ion Coulomb crystals as a media for multi-mode light-matter interfacing, a precise measurement of the coherent coupling rates for both modes was obtained by again measuring the broadening of the probe pulse transmission signal but this time for a series of detunings of the probe pulse from atomic resonance, $\Delta$. To obtain high optical densities, in this experiment, a crystal with a larger density ($\rho=5.4\times 10^{8}$ cm$^{-3}$) and almost twice the length ($2L=1.2$ mm) of the crystals used in the previous experiment was used. 
In Fig.~\ref{fig4}, a projection image of this crystal is shown together with images indicating the position of the TEM$_{00}$ and TEM$_{10}$ modes applied in the experiments. The results are presented in Fig.~\ref{fig5} together with Lorentzian fits to the data based on Eq.~\ref{kappa}.
With $G_{mn}$ and $\gamma$ as free fitting parameters, we obtain
($G_{00}=(2\pi)11.6\pm 0.1$ MHz, $\gamma=(2\pi)11.3\pm0.3$ MHz) and
($G_{10}=(2\pi)11.5\pm 0.1$ MHz, $\gamma=(2\pi)11.4\pm0.3$ MHz) for the TEM$_{00}$ and TEM$_{10}$ mode, respectively. The collective coupling rates, $G_{00}$ and $G_{10}$, are strikingly identical. 
The obtained $G_{00}$ and $G_{10}$ correspond to optical depths $G^2/\kappa\gamma\simeq 4.5$, but could easily be made larger as is presented in Ref.~\cite{herskind09NatPhys}.
Furthermore, an equal coupling strength was also observed for
the TEM$_{10}$ and the TEM$_{01}$ mode, indicating that ion Coulomb crystals open up for e.g. the possibility of storing photonic qubits encoded in a spatial basis spanned by these two modes.

In conclusion we have shown that ion Coulomb crystals are an ideal system for obtaining high and uniform coupling strengths to the field of an optical cavity independent of the transverse mode.  The achieved excellent agreement between experimental data and theoretical predictions confirm that the presence of the rf-trapping fields does not significantly alter the coupling strengths even for ions far from the trap axis. These results combined with previously measured long collective Zeeman sub-state coherence times~\cite{herskind09NatPhys} clearly indicate that ion Coulomb crystals should be able to serve as near-ideal media for high-fidelity multi-mode quantum information processing and communication devices.

We acknowledge financial support from the Carlsberg Foundation and
the Danish Natural Science Research Council through the ESF EuroQUAM
project CMMC.\\

\bibliographystyle{apsrev}
%\bibliography{../../whole_ions}

\end{document}